\documentclass[preprint,copyright,creativecommons]{eptcs}
\providecommand{\event}{SOS 2007} 

\usepackage{iftex}
\usepackage[most]{tcolorbox}
\usepackage{xcolor}
\usepackage{pifont}
\usepackage{tabularx}

\ifpdf
  \usepackage{underscore}         
  \usepackage[T1]{fontenc}        
\else
  \usepackage{breakurl}           
\fi

\usepackage{amssymb, amsthm, amsmath}
\theoremstyle{definition}
\newtheorem{definition}{Definition}[section]

\newtheorem{principle}{Principle}

\newcommand{\tick}{\textcolor{green!60!black}{\ding{51}}}
\newcommand{\cross}{\textcolor{red}{\ding{55}}}

\usepackage{tikzit}
\input{quantum.tikzdefs}
\tikzstyle{gate}=[shape=rectangle, text height=1.5ex, text depth=0.25ex, yshift=-0.5mm, fill=white, draw=black, minimum height=5mm, minimum width=5mm, font={\small}, tikzit category=circuit]
\tikzstyle{big gate}=[shape=rectangle, text height=1.5ex, text depth=0.25ex, yshift=-0.5mm, fill=white, draw=black, minimum height=10mm, minimum width=5mm, font={\small}, tikzit category=circuit]
\tikzstyle{Z dot}=[inner sep=0mm, minimum size=2mm, shape=circle, draw=black, fill=zxgreen, tikzit fill={rgb,255: red,221; green,255; blue,221}, tikzit category=zx]
\tikzstyle{Z bold dot}=[inner sep=0mm, minimum size=2mm, shape=circle, draw=black, fill=zxgreen, tikzit fill={rgb,255: red,221; green,255; blue,221}, line width=1.2pt, tikzit category=zx]
\tikzstyle{Z phase dot}=[minimum size=5mm, font={\footnotesize\boldmath}, shape=rectangle, rounded corners=2mm, inner sep=0.2mm, outer sep=-2mm, scale=0.8, tikzit shape=circle, draw=black, fill=zxgreen, tikzit fill={rgb,255: red,221; green,255; blue,221}, tikzit draw=blue, tikzit category=zx]
\tikzstyle{Z tiny dot}=[inner sep=0.2mm, font={\footnotesize\boldmath}, minimum size=1mm, shape=circle, draw=black, fill=zxgreen, tikzit fill={rgb,255: red,221; green,255; blue,221}]
\tikzstyle{X dot}=[Z dot, shape=circle, draw=black, fill=zxred, tikzit fill={rgb,255: red,255; green,136; blue,136}, tikzit category=zx]
\tikzstyle{X bold dot}=[inner sep=0mm, minimum size=2mm, shape=circle, draw=black, fill=zxred, tikzit fill={rgb,255: red,255; green,136; blue,136}, line width=1.2pt, tikzit category=zx]
\tikzstyle{X phase dot}=[Z phase dot, tikzit shape=circle, tikzit draw=blue, fill=zxred, tikzit fill={rgb,255: red,255; green,136; blue,136}, font={\footnotesize\boldmath}, tikzit category=zx]
\tikzstyle{X tiny dot}=[inner sep=0.2mm, font={\footnotesize\boldmath}, minimum size=1mm, shape=circle, draw=black, fill=zxred, tikzit fill={rgb,255: red,255; green,136; blue,136}]
\tikzstyle{hadamard}=[fill=yellow, draw=black, shape=rectangle, inner sep=0.6mm, minimum height=1.5mm, minimum width=1.5mm, tikzit category=zx]
\tikzstyle{paulibox}=[fill={rgb,255: red,221; green,221; blue,255}, draw=black, shape=rectangle, inner sep=0.6mm, minimum height=5mm, minimum width=5mm, font={\footnotesize}, text height=1.5ex, text depth=0.25ex, tikzit category=zx]
\tikzstyle{vertex}=[inner sep=0.2mm, minimum size=1mm, shape=circle, draw=black, fill=black, tikzit category=misc]
\tikzstyle{vertex set}=[inner sep=0.2mm, minimum size=1mm, shape=circle, draw=black, fill=white, font={\footnotesize\boldmath}, tikzit category=misc]
\tikzstyle{small black dot}=[fill=black, draw=black, shape=circle, inner sep=0pt, minimum width=1.2mm, tikzit category=circuit]
\tikzstyle{cnot ctrl}=[fill=black, draw=black, shape=circle, inner sep=0pt, minimum width=1.2mm, tikzit category=circuit]
\tikzstyle{cnot targ}=[fill=white, draw=white, shape=circle, tikzit category=circuit, label={center:$\oplus$}, inner sep=0pt, minimum width=2.1mm, tikzit fill={rgb,255: red,102; green,204; blue,255}, tikzit draw=black]
\tikzstyle{ket}=[fill=white, draw=black, shape=regular polygon, regular polygon sides=3, regular polygon rotate=-30, scale=0.7, inner sep=1pt, tikzit category=circuit, tikzit shape=rectangle, tikzit fill=green]
\tikzstyle{bra}=[fill=white, draw=black, shape=regular polygon, regular polygon sides=3, regular polygon rotate=30, scale=0.7, inner sep=1pt, tikzit category=circuit, tikzit shape=rectangle, tikzit fill=red]
\tikzstyle{scalar}=[shape=rectangle, text height=1.5ex, text depth=0.25ex, yshift=-0.5mm, fill=white, draw=black, minimum height=5mm, minimum width=5mm, font={\small}]
\tikzstyle{clabel}=[fill=white, draw=none, shape=rectangle, tikzit fill={rgb,255: red,56; green,255; blue,242}, font={\footnotesize}, inner sep=1pt, tikzit category=labels]
\tikzstyle{empty diagram}=[draw=gray!40!white, dashed, shape=rectangle, minimum width=1cm, minimum height=1cm, tikzit category=misc]
\tikzstyle{amap}=[fill=white, draw=black, shape=NEbox, tikzit category=asymmetric, tikzit fill=yellow, tikzit shape=rectangle]
\tikzstyle{amap conj}=[fill=white, draw=black, shape=NWbox, tikzit category=asymmetric, tikzit fill=green, tikzit shape=rectangle]
\tikzstyle{amap adj}=[fill=white, draw=black, shape=SEbox, tikzit category=asymmetric, tikzit fill=red, tikzit shape=rectangle]
\tikzstyle{amap trans}=[fill=white, draw=black, shape=SWbox, tikzit category=asymmetric, tikzit fill=orange, tikzit shape=rectangle]
\tikzstyle{astate}=[fill=white, draw=black, shape=NEtriangle, tikzit category=asymmetric, tikzit shape=circle, tikzit fill=yellow]
\tikzstyle{astate conj}=[fill=white, draw=black, shape=NWtriangle, tikzit category=asymmetric, tikzit shape=circle, tikzit fill=green]
\tikzstyle{astate adj}=[fill=white, draw=black, shape=SEtriangle, tikzit category=asymmetric, tikzit shape=circle, tikzit fill=red]
\tikzstyle{astate trans}=[fill=white, draw=black, shape=SWtriangle, tikzit category=asymmetric, tikzit shape=circle, tikzit fill=orange]
\tikzstyle{box}=[shape=rectangle, text height=1.5ex, text depth=0.25ex, yshift=-0.5mm, fill=white, draw=black, minimum height=5mm, minimum width=5mm, font={\small}]
\tikzstyle{medium box}=[shape=rectangle, text height=1.5ex, text depth=0.25ex, yshift=-0.5mm, fill=white, draw=black, minimum height=10mm, minimum width=5mm, font={\small}]
\tikzstyle{simple}=[-]
\tikzstyle{hadamard edge}=[-, dashed, dash pattern=on 2pt off 0.5pt, thick, draw={rgb,255: red,68; green,136; blue,255}]
\tikzstyle{box edge}=[-, dashed, dash pattern=on 2pt off 0.5pt, thick, draw={rgb,255: red,203; green,192; blue,225}]
\tikzstyle{brace edge}=[-, tikzit draw=blue, decorate, decoration={brace,amplitude=1mm,raise=-1mm}]
\tikzstyle{diredge}=[->]
\tikzstyle{double edge}=[-, double, shorten <=-1mm, shorten >=-1mm, double distance=2pt]
\tikzstyle{gray edge}=[-, gray!60!white]
\tikzstyle{pointer edge}=[->, very thick, gray]
\tikzstyle{boldedge}=[-, line width=1.2pt, shorten <=-0.17mm, shorten >=-0.17mm]
\tikzstyle{bidir edge}=[<->, very thick, draw={rgb,255: red,191; green,191; blue,191}]
\tikzstyle{surface X}=[-, tikzit fill=red, fill=zxred]
\tikzstyle{surface Z}=[-, tikzit fill=green, fill=zxgreen]
\tikzstyle{X Web}=[-, preaction={line width=1.5mm, draw={rgb,255: red,255; green,150; blue,150}, dashed, dash pattern=on 2.25pt off 0.5pt}, shorten <=-0.25mm, shorten >=-0.25mm, tikzit draw=red]
\tikzstyle{Y Web}=[-, preaction={line width=1.5mm, draw={rgb,255: red,150; green,150; blue,255}, dashed, dash pattern=on 1.2pt off 0.5pt}, shorten <=-0.25mm, shorten >=-0.25mm, tikzit draw=blue]
\tikzstyle{Z Web}=[-, preaction={line width=1.5mm, draw={rgb,255: red,120; green,200; blue,120}}, shorten <=-0.25mm, shorten >=-0.25mm, tikzit draw={rgb,255: red,0; green,141; blue,0}]
\tikzstyle{light-fill}=[-, fill={rgb,255: red,230; green,230; blue,230}, tikzit fill={rgb,255: red,191; green,191; blue,191}, draw=none, tikzit draw={rgb,255: red,191; green,191; blue,191}]
\tikzstyle{dashed edge}=[-, dashed, dash pattern=on 2pt off 0.5pt, draw=black]
\tikzstyle{light dashed edge}=[-, dashed, dash pattern=on 2pt off 0.5pt, draw={rgb,255: red,191; green,191; blue,191}]

\title{ABSTRACTS:\\Amsterdam Benchmark Suite for the Time and Resource Analysis of Clifford+T Simulators}
\author{Matthew Sutcliffe \qquad\qquad John van de Wetering
\institute{University of Amsterdam, QuSoft\\Amsterdam, The Netherlands}
\email{\quad m.j.sutcliffe@uva.nl \quad\qquad j.m.m.vandewetering@uva.nl}
}
\def\titlerunning{ABSTRACTS}
\def\authorrunning{M. Sutcliffe \& J. van de Wetering}
\begin{document}
\maketitle

\begin{abstract}
Recent years have seen a rapid growth in literature presenting new methods for simulating non-Clifford quantum circuits with classical hardware. These methods span a range of approaches, including stabiliser decomposition and tensor contraction techniques, varying in efficiency depending on circuit class, depth, non-Clifford gate count, and other metrics. A notable limitation of this literature is the lack of a standardised approach to benchmarking, with each new paper outlining its own specification, simulating its own set of circuits on the authors' own hardware. This paper seeks to address this issue by presenting a standardised and canonical benchmark suite and infrastructure for quantifying the efficiency of non-Clifford classical simulators, with a consistent dataset of circuits and providing consistent (virtual) hardware, thereby enabling a fair comparison of results.
\end{abstract}

\section{Introduction}

A core area of research in quantum computing is \textit{classical simulation}: the task of computing quantum circuits with classical hardware. This is vital for verifying quantum algorithms and computers, as well as understanding and quantifying the quantum advantage. Many approaches have been taken in solving this problem, with methods based on state vector simulation \cite{qiskit2024,suzuki2021qulacs}, tensor contraction \cite{gray2021hyper,vincent2022jet}, stabiliser decomposition \cite{kissinger2022simulating,sutcliffe-thesis}, and beyond. In the general `\textit{non-Clifford}' case, every known method scales exponentially against one circuit metric or another, such as qubit count, non-Clifford gate count, or treewidth. Many new methods, techniques, and heuristics have been presented in recent years, demonstrating reduced rates of exponential scaling and hence more efficient classical simulation. This in turn enables larger and more complex quantum circuits to be simulable classically and raises the bar quantum computers and algorithms must meet to achieve quantum advantage \cite{huang2020classical,pan2021simulating,manabe2026classical}.

With the literature on this topic growing rapidly, a notable problem has arisen, namely the lack of a standardised approach to benchmarking. In general, every new paper presenting a new simulation technique has outlined its own set of benchmark circuits and run experiments on the authors' own hardware, following their own methodology, often benchmarked alongside any one among many existing methods for comparison. While this was sufficient when the literature was small and few competing methods existed, it has since made fair comparisons of simulation techniques difficult.

\begin{figure}[!htb]
    \centering
    \includegraphics[width=\linewidth]{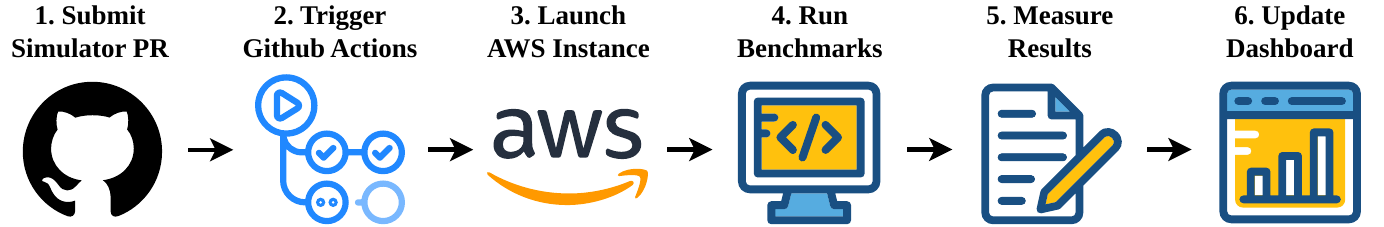}
    \caption{Overview of the automated ABSTRACTS benchmarking procedure, triggered when a new simulator or circuit (class) submission is accepted into the repository.}
    \label{fig:architecture}
\end{figure}

To address this issue, we present the \textit{Amsterdam Benchmark Suite for the Time and Resource Analysis of Clifford+T Simulators} (\textit{ABSTRACTS}), or simply the \textit{Amsterdam Benchmark Suite}, as a standardised and canonical benchmark dataset and framework for assessing classical simulators of non-Clifford quantum circuits, with an automated execution pipeline as illustrated in Figure~\ref{fig:architecture}. This provides consistency in the benchmark setup, methodology, circuits, and hardware spec, as well as a record of results for many existing simulators.

\begin{figure}[!htb]
    \centering
    \includegraphics[width=\linewidth]{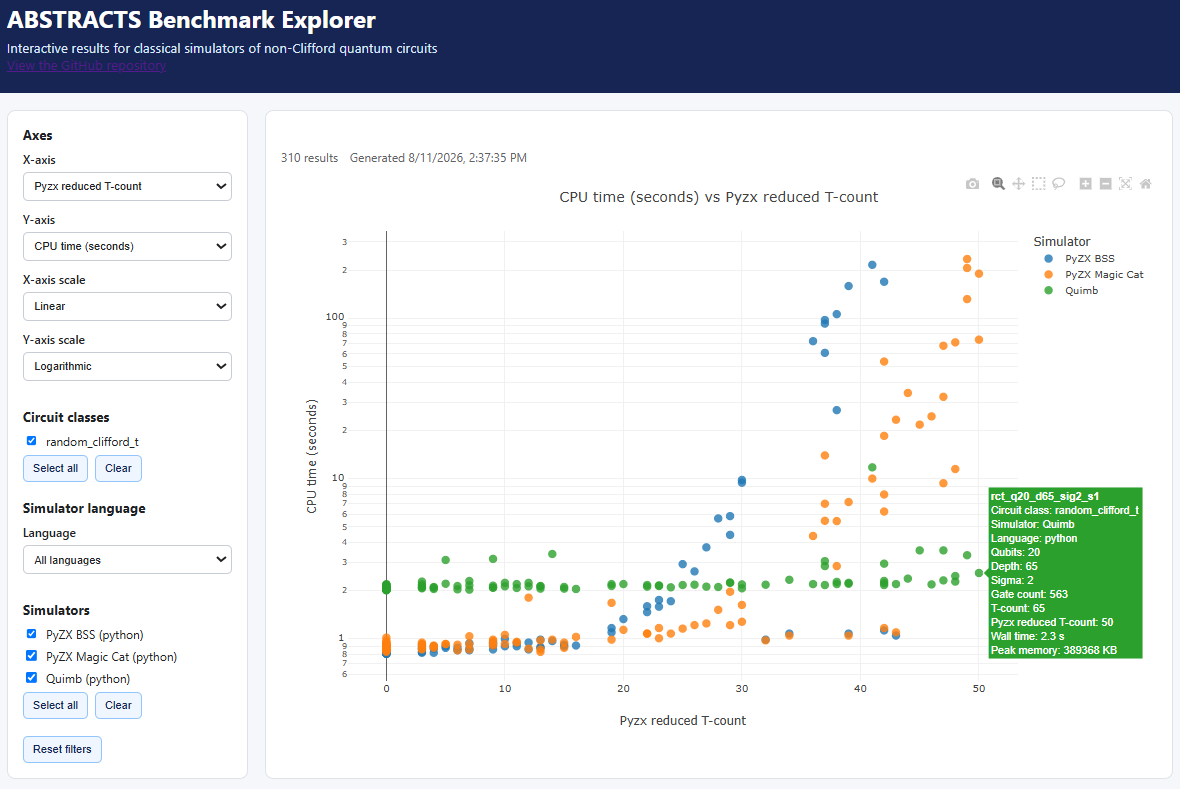}
    \caption{The ABSTRACTS `\textit{dashboard}', hosted at {\url{https://mjsutcliffe99.github.io/ABSTRACTS/}}, for visualisation and analysis of its benchmark results.}
    \label{fig:dashboard}
\end{figure}

In Figure~\ref{fig:dashboard} we show the output of this process, namely a dashboard that allows for easy comparison among simulation methods for different classes of circuits and across a range of metrics.

\begin{figure}[h]
    \centering
    \includegraphics[width=\linewidth]{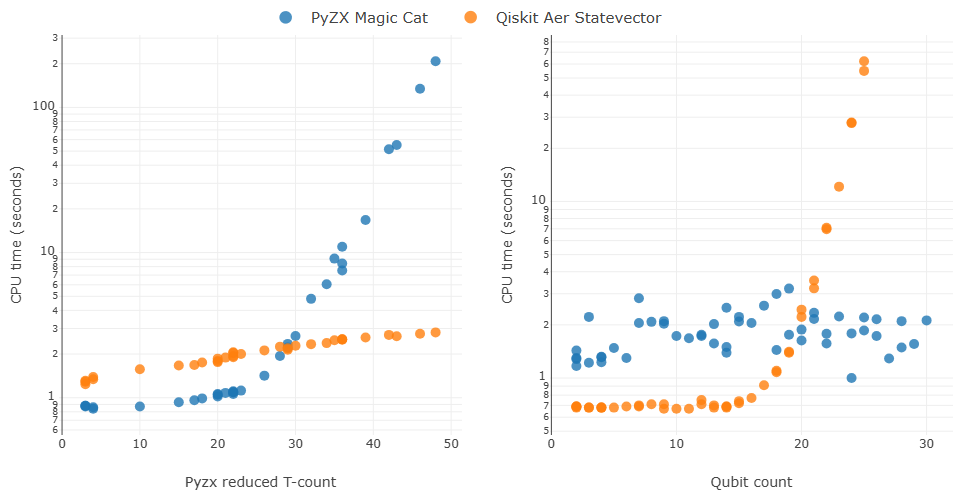}
    \caption{Two plots generated via the ABSTRACTS dashboard, comparing the CPU time for simulating random Clifford+T circuits with the PyZX `Magic Cat' \cite{kissinger2022sim2} stabiliser decomposition strategy and the Qiskit Aer Statevector simulator \cite{qiskit2024}. The random Clifford+T circuits are of (left) a fixed qubit count of 20, and (right) a fixed T-count (after plugged ZX-calculus simplification) of 30, respectively. Observe the difference in exponential scaling across two different metrics, showing that neither method is better than the other in all cases.}
    \label{fig:plots}
\end{figure}

Appropriately comparing the performance of simulators requires nuance, as there are myriad approaches to classical simulation, optimising against a range of different metrics. For instance, one simulator might perform very well on a particular circuit class but poorly on another, or one may scale well as T-count grows but poorly against treewidth, with the opposite being true for an alternative simulator. Similarly, some simulators may trade speed for memory usage or vice versa, or one may incur a slow initial overhead but manage especially high-speed subsequent simulation sampling. See, for instance, Figure~\ref{fig:plots}.

Consequently, in general, it is not meaningful to speak of one simulator being `better' than another in a simplistic global sense. Rather, the nuances of how each simulator behaves and scales against different circuit classes and metrics is integral to the analysis of comparing their performances. With this principle at the forefront of this project, the goal of ABSTRACTS is not to provide a definitive linear leaderboard of classical simulators but rather to collate and present thorough data about how each simulator performs to facilitate such a nuanced analysis.

The project aims to observe and, where appropriate, quantify how different simulation methods behave under various conditions and how they scale against different metrics. It is often, therefore, less interested in the \textit{specific} numerical results (of runtime or memory usage, etc.) than in the relative trends and scaling patterns of the simulators. As such, even simulators of wildly different implementation details, such as a relatively slow Python prototype of one method versus a high-performance multithreaded Rust implementation of another, can still be of interest to compare (though this context should of course always be taken into account when performing any analysis in comparing such simulators). Nevertheless, for those interested in comparing competitive, high-performance simulators, the consistent benchmark suite and infrastructure ensure the project is likewise suitable for this secondary purpose.

Researchers and developers creating new classical simulation techniques and software packages are encouraged to submit their simulators for official ABSTRACTS benchmarking at:

\par\vspace{0.5em}
\noindent\makebox[\textwidth][c]{\url{https://github.com/mjsutcliffe99/ABSTRACTS}}
\par\vspace{0.5em}

The results may then be visualised and analysed on the ABSTRACTS dashboard at:

\par\vspace{0.5em}
\noindent\makebox[\textwidth][c]{\url{https://mjsutcliffe99.github.io/ABSTRACTS/}}
\par\vspace{0.5em}

\begin{figure*}[!htbp]
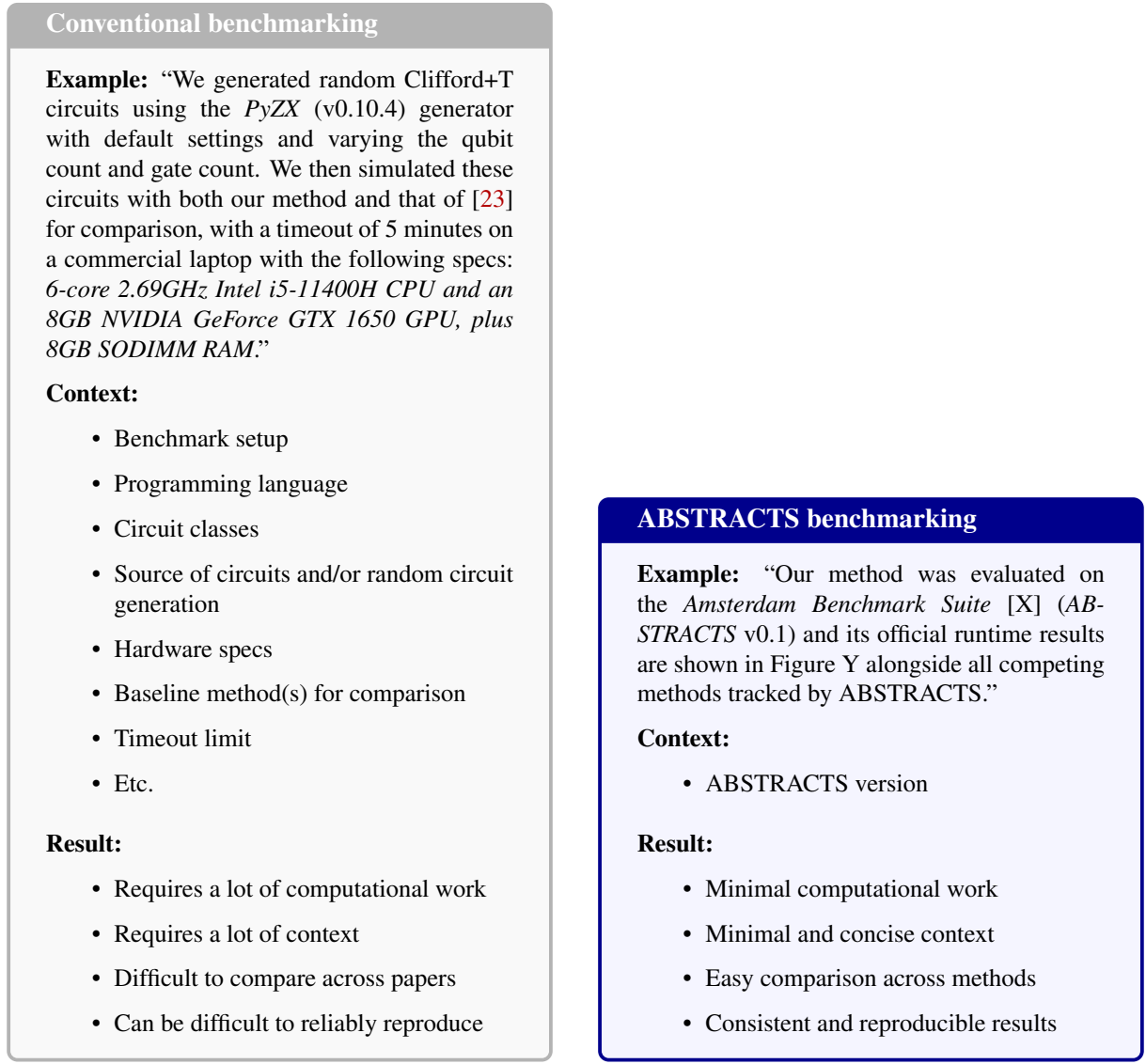

    \centering
    \begin{minipage}[t]{0.48\textwidth}
        \begin{tcolorbox}[
            title={Conventional benchmarking},
            colback=gray!5,
            colframe=gray!60,
            fonttitle=\bfseries
        ]
        \small

        \textbf{Example:} ``We generated random Clifford+T circuits using the \textit{PyZX} (v0.10.4) generator with default settings and varying the qubit count and gate count. We then simulated these circuits with both our method and that of \cite{kissinger2022sim2} for comparison, with a timeout of 5 minutes on a commercial laptop with the following specs: \textit{6-core 2.69GHz Intel i5-11400H CPU and an 8GB NVIDIA GeForce GTX 1650 GPU, plus 8GB SODIMM RAM}.''

        \medskip
        \textbf{Context:}
        \begin{itemize}
            \item Benchmark setup
            \item Programming language
            \item Circuit classes
            \item Source of circuits and/or random circuit generation
            \item Hardware specs
            \item Baseline method(s) for comparison
            \item Timeout limit
            \item Etc.
        \end{itemize}

        \medskip
        \textbf{Result:}
        \begin{itemize}
            \item Requires a lot of computational work
            \item Requires a lot of context
            \item Difficult to compare across papers
            \item Can be difficult to reliably reproduce
        \end{itemize}
        \end{tcolorbox}
    \end{minipage}
    \hfill
    \begin{minipage}[t]{0.48\textwidth}
        \begin{tcolorbox}[
            title={ABSTRACTS benchmarking},
            colback=blue!4,
            colframe=blue!55!black,
            fonttitle=\bfseries
        ]
        \small
        \textbf{Example:} ``Our method was evaluated on the \textit{Amsterdam Benchmark Suite} [X] (\textit{ABSTRACTS} v0.1) and its official runtime results are shown in Figure Y alongside all competing methods tracked by ABSTRACTS.''
        
        \medskip
        \textbf{Context:}
        \begin{itemize}
            \item ABSTRACTS version
        \end{itemize}

        \medskip
        \textbf{Result:}
        \begin{itemize}
            \item Minimal computational work
            \item Minimal and concise context
            \item Easy comparison across methods
            \item Consistent and reproducible results
        \end{itemize}
        \end{tcolorbox}
    \end{minipage}
    \caption{A comparison of (left) the heavily contextual and verbose benchmarking summary typical of papers in the literature to date, and (right) a corresponding simple and concise summary facilitated by benchmarking with ABSTRACTS. In the latter case, the relevant computational details may be found on the GitHub repository from the ABSTRACTS version alone, rather than needing to be contextualised within each paper.}
    \label{fig:abs-reporting}
\end{figure*}

For greater context and reproducibility, it is recommended to include the version number when referencing results presented in the benchmark suite, e.g. ``\textit{benchmarked on ABSTRACTS v0.1}'', as exemplified in Figure \ref{fig:abs-reporting}.

We emphasise that the principle contribution of this paper is presenting the benchmarking infrastructure provided by ABSTRACTS, rather than an exhaustive benchmark study of classical simulation techniques. The benchmarking results included serve primarily to demonstrate the framework, with a more extensive benchmarking and analysis of simulators intended for future work. In short, we present this work as a `tool paper' rather than a `benchmark paper'.

\section{Background on classical simulation}

The field of classical simulation of quantum computations is vast, and there are many (closely) related problems that different sub-problems might be tackling. To make clear which problem we are solving with the methods we benchmark, we give here an overview of the main classes of simulation problems.

\subsection{Strong vs weak simulation}

\begin{definition}[Weak Simulation]
    Let $C$ be a quantum circuit, and write
    $$P(x_1,\ldots,x_n) = \lvert \bra{x_1\cdots x_n}U\ket{0\cdots 0}\rvert^2$$ 
    for the probabilities of observing the outcome $\ket{x_1\cdots x_n}$ when applying $C$ to the input state $\ket{0\cdots 0}$.
    We then say a probabilistic algorithm \emph{weakly simulates} $C$ when it produces as output bit strings $\vec y \in \{0,1\}^n$ according to a distribution `suitably close' to $P$ \cite{KissingerWetering2024Book}.
\end{definition}
An algorithm that weakly simulates a quantum computation produces the same kind of output as what you would actually get from running the computation on real quantum hardware, namely a set of measurement outcomes whose likelihood depends on the specifics of the circuit. We require the output to be given with a probability that is `suitably close' to the target distribution. What this means exactly might differ by the model considered. We could for instance consider \emph{exact} weak simulation, where the samples must be statistically indistinguishable from the true distribution. The more common notion is where we allow a small additive error, which matches the reality where we would only ever be able to compare a polynomial number of samples from the true quantum circuit and the simulator (concretely, we could require that the Total Variation distance between the distributions is smaller than some small $\varepsilon$).

There are a number of weak simulation methods, but we will explain these after we've introduced the related notion of strong simulation.

\begin{definition}[Strong Simulation]
  Let $C$ be a quantum circuit, and let $P(x_1,\ldots,x_n)$ be its associated probability distribution as above. Then we say an algorithm \emph{strongly simulates} $C$ when it can calculate (or closely approximate up to multiplicative error) any marginal probability of $P$ \cite{KissingerWetering2024Book}.
\end{definition}
So whereas the output of a weak simulator is a sample from the circuit, the output of a strong simulator is a number representing a probability. As the name implies, using strong simulation we can do weak simulation. The most well-known way to do this is using the qubit-by-qubit method, where we first calculate the marginal probability of the first qubit being 1, and then sampling from it with this probability. Then we calculate the marginal probability of the second qubit, conditional on the first qubit being the value we just sampled. Iterating this for all qubits gives us one full sample of all the qubit outcomes, at the cost of running our strong simulation oracle $n$ times. We can also use the gate-by-gate method~\cite{bravyi2022measurement} (or, similarly, the more recent T-by-T method~\cite{Koch_sampling}), which is a bit more intricate but involves the calculation of a number of probabilities that scales with the number of gates in the circuit, not the number of qubits, but doesn't require any marginal probability calculations (which, as we will see later, is beneficial for certain protocols).

As the name also implies, we cannot perform strong simulation using a weak simulator. The thing we would want to try is to generate many samples in order to approximate a marginal probability. However, it could be that a marginal probability is exponentially small (this is in fact the case for many quantum algorithms), and hence we would need exponentially many samples to get a multiplicatively accurate estimate (this is also why the requirement of the error being multiplicative instead of additive is important in the definition of strong simulation). In particular, a quantum computer cannot efficiently perform strong simulation (unless certain complexity classes collapse). Formally, we have that weak simulation is BQP-complete, and hence exactly captures the power of a quantum computer, while strong simulation is \#P-hard (and given the right formal definitions is in fact \#P-complete), and hence is believed to be much harder. One way to see this is that for strong simulation it doesn't matter whether we are simulating a regular quantum computation or a post-selected quantum computation, and we know that allowing for post-selecting greatly boosts the power of a quantum computer~\cite{aaronson}.

While it is true that strong simulation is the harder task, most techniques we have for weak simulation go via strong simulation and then use either the qubit-by-qubit or gate-by-gate method. The exceptions are the \emph{quasi-probabilistic} simulators such as those based on the Wigner distribution \cite{pashayan2015quasiprobabilities} or stabiliser extent \cite{bravyi2019stabilizer}. In these simulators, we write our hard to simulate computation as a linear combination of (potentially exponentially many) simple to simulate computations. By then Monte-Carlo sampling from these easy to simulate computations we get an unbiased estimate of the observable we care about, though at the cost of an exponentially increased variance, so that we require more samples to get the desired precision.

As most simulation techniques do strong simulation, we focus on this task in this paper and leave benchmarking weak simulators for future work.

\subsection{Approaches to strong simulation}

The most basic way to do strong simulation is state-vector simulation \cite{qiskit2024,suzuki2021qulacs}. There we represent the starting quantum state as an exponentially long vector which is updated by every quantum gate that is applied to it. Any marginal probability can then be easily calculated from this vector directly. This has the obvious drawback of taking exponential memory and time in the number of qubits.

A more sophisticated method is to use tensor networks \cite{gray2018quimb,gray2021hyper,vincent2022jet}. We represent the marginal probability or amplitude as a closed network of tensors (closed meaning there are no uncontracted indices) and then try to find an optimal sequence of contractions that keeps the dimensions of the intermediate tensors as small as possible. The standard way to do this scales exponentially with the \emph{treewidth} of the tensor network, which might be significantly smaller than the number of qubits. Approximate contraction can also be done by using singular value decompositions to throw away small eigenvalues of intermediate tensors.

Another notable family of techniques are those based on \emph{stabiliser decompositions}, wherein the simulation scales exponentially in the non-Clifford gate count of the circuit. Fully Clifford circuits (those whose only gates are Hadamard, $S$ and CNOT) can be simulated efficiently by the Gottesman-Knill theorem \cite{gottesman1999heisenberg,aaronson2004improved}. By decomposing every non-Clifford gate, like the Toffoli or $T$ gate, as a sum of Clifford gates, we can represent any quantum computation as an (exponentially sized) sum of efficient-to-simulate Clifford computations. The main benefit of these techniques is that they scale well with number of qubits and number of Clifford gates and hence are suitable for `near-Clifford' computations that for instance cover many protocols in fault-tolerant quantum computing.

The exponential scaling of stabiliser rank techniques depends on how many terms are needed to represent the non-Clifford resources used in the computation. For a Clifford+$T$ circuit with $k$ $T$ gates, we write the number of terms as $2^{\alpha k}$ where $\alpha\in \mathbb{R}_{>0}$ represents the factor that determines the exponential growth. Different decomposition strategies have a different $\alpha$, and certain techniques might have a better (lower) `effective' $\alpha$ for certain classes of circuits where many terms can cancel out. The best-known $\alpha$ that holds for arbitrary circuits is $\alpha\approx 0.396$ \cite{kissinger2022sim2,qassim2021improved}.

\section{Benchmark Architecture}

\subsection{Infrastructure}

The core pillars of the ABSTRACTS infrastructure consist of:
\begin{itemize}
    \item circuits, 
    \item simulators,
    \item the cloud compute instance,
    \item results, and
    \item the dashboard.
\end{itemize}

\begin{figure}
    \centering
    \includegraphics[width=\linewidth]{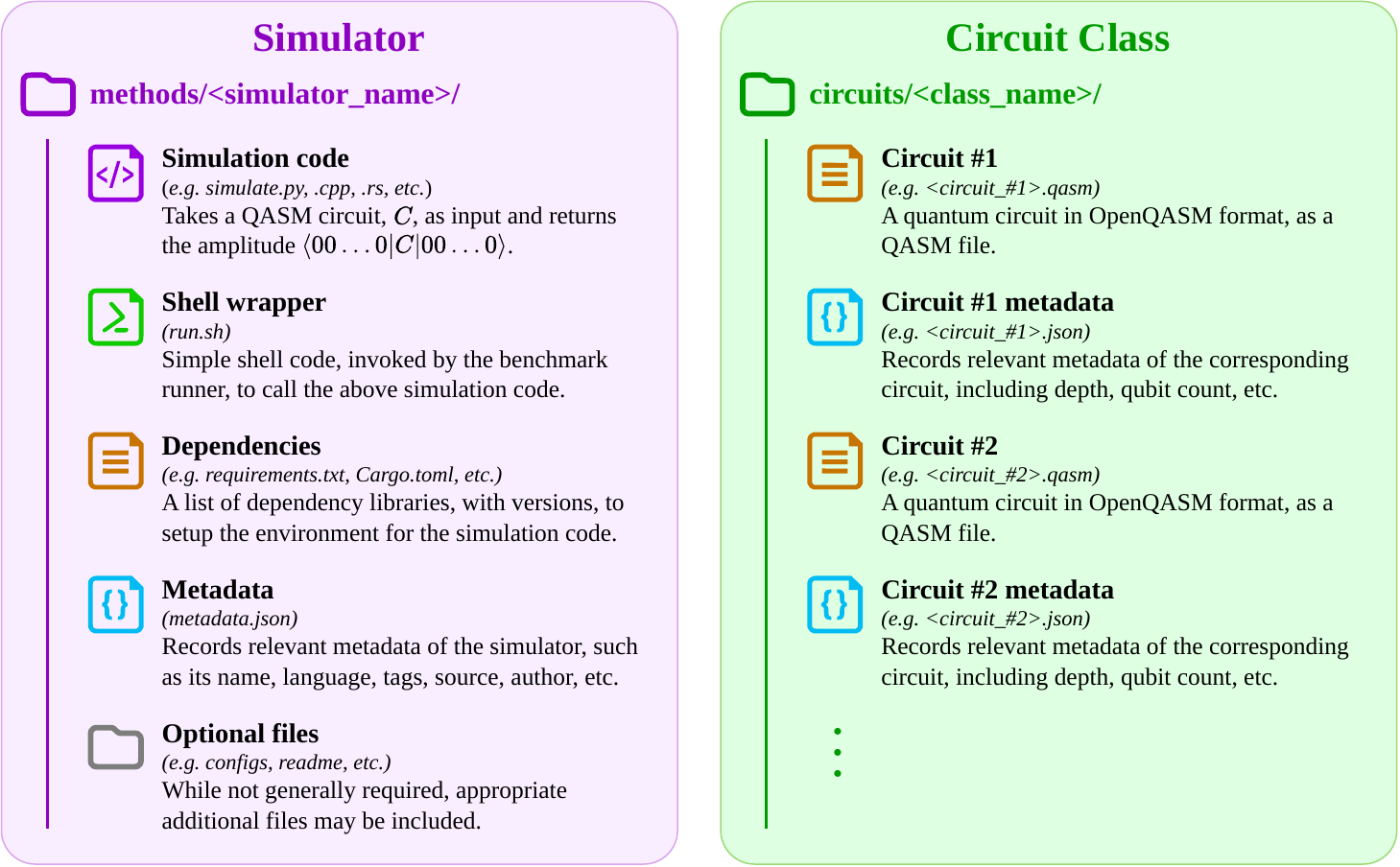}
    \caption{The file structures of (left) a simulator and (right) a circuit class in the ABSTRACTS repo.}
    \label{fig:sim-circ}
\end{figure}

In particular, ABSTRACTS includes a large set of circuits, organised into subsets by circuit class, as well as a growing set of strong classical simulators, and is designed to be extensible on both accounts, welcoming new submissions of circuits and simulators to be included in the benchmarking suite. More specifically, Figure~\ref{fig:sim-circ} shows what constitutes a simulator and a circuit class, and examples can be found on the GitHub repo \cite{abstracts_repo}.

\begin{figure}
    \centering
    \includegraphics[width=\linewidth]{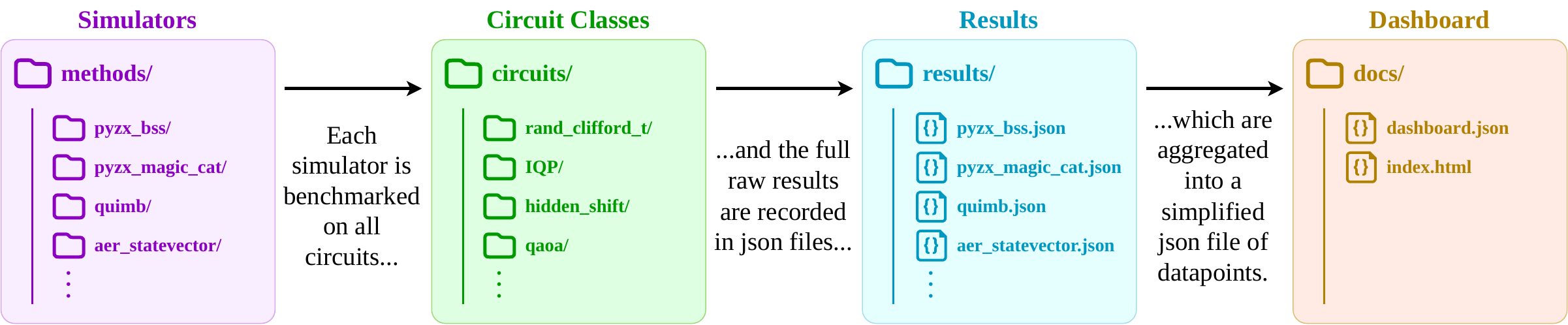}
    \caption{An illustration of the repository structure and how its various components interact.}
    \label{fig:repo}
\end{figure}

Fully benchmarking a simulator consists of strongly simulating every circuit with its method on the cloud compute instance and recording the tracked results per circuit, including the CPU time, wall time, peak memory usage, precision, etc. As such, every benchmarked simulator produces a \textit{json} file tracking how effectively and efficiently it managed to simulate each circuit. Lastly, whenever a new benchmark is executed, an automated script aggregates the raw data from every simulator into a single simplified results file used to generate the plots on the interactive webpage, or `\textit{dashboard}'. These relationships among the infrastructure components are summarised in Figure~\ref{fig:repo}.

While details of the execution pipeline are liable to change across versions, and up to date information may be found on the GitHub repository \cite{abstracts_repo}, the following specifications hold true for ABSTRACTS v0.1:
\begin{itemize}
    \item The virtual hardware on which all benchmarks are executed is the \textit{t3.small} Elastic Compute Cloud (EC2), provided by Amazon Web Services, consisting of 2 vCPUs and 2GiB of RAM.
    \item The EC2 instance runs \textit{Ubuntu 26.04 LTS (Resolute Raccoon)} as its operating system.
    \item The benchmark includes a local 5 minute timeout per circuit, and a global 3 hour timeout per simulator.
    \item In benchmarking a simulator, a successfully simulated circuit will record the following raw results: computed amplitude, wall time, CPU time, and peak memory usage.
    \item In benchmarking a simulator, an unsuccessfully simulated circuit should track one of the following `failed' statuses: exceeded circuit timeout, exceeded global benchmark timeout, exceeded memory limit\footnote{An incorrect or imprecise amplitude may be tracked or filtered out from the \textit{aggregated} data (and hence the corresponding dashboard plots) by specifying an acceptable margin of error. In this way, approximate simulators are not marked as failed in the raw data, nor is an acceptable margin of error hard-coded.}.
\end{itemize}

Since the aim of ABSTRACTS is not to showcase the frontier of what is classically simulable, but rather to demonstrate how various simulation techniques scale across circuit metrics, these relatively restrictive hardware and timeout specs are sufficient.

\subsection{Design principles}

As mentioned in the introduction, the aim of ABSTRACTS is not to provide a linear leaderboard of which simulator is the best in all circumstances, but rather to quantify scalings in different metrics and to make it possible to compare wildly different methods on equal footing.

In addition to the above mission statement, we adhere to the following four fundamental design principles:

\begin{itemize}
    \item \textbf{Consistency} allows for a fair comparison of the performance results of the simulators.
    \item \textbf{Reproducibility} ensures the results are consistent and reliable over time.
    \item \textbf{Extensibility} ensures new simulation techniques and circuit classes can be easily added for benchmarking in the future, allowing ABSTRACTS to continue to track the state of the art as it develops over time.
    \item \textbf{Transparency} ensures the results produced by ABSTRACTS can be trusted and relied upon as honest and fair reflections of the performances of the simulators.
\end{itemize}

The core premise of ABSTRACTS is to provide a consistent benchmark suite and infrastructure to enable fair and easy comparisons of non-Clifford classical simulators over a range of metrics. Towards this aim, per ABSTRACTS version, we maintain a consistent dataset of circuits against which simulators are benchmarked and execute all benchmarks on the same (virtual) hardware.

However, the collection of circuits and the hardware spec are liable to undergo expansion and changes from one ABSTRACTS version to the next as needed, such as to include new circuit classes or to migrate to a new cloud compute instance should the current one be discontinued or superseded. Nevertheless, version history will be appropriately maintained and documented and consistency maintained \textit{per version} as detailed in Principle~\ref{princ:consistency}.

\hfill\begin{principle}[Consistency]
\label{princ:consistency}
Per ABSTRACTS version, consistency is maintained across the following:
\begin{itemize}
    \item Benchmark dataset of circuits (including metadata).
    \item Virtual hardware (i.e. cloud compute instance).
    \item Execution environment (e.g. operating system, runtime, etc.).
    \item Execution protocol (e.g. results schema, timeout configuration, etc.).
\end{itemize}
An update to any of these itemised points (such as adding new circuits to the repo or migrating the cloud compute instance) will constitute a new ABSTRACTS version and trigger a rerun of all benchmarks. As such, \textbf{an ABSTRACTS version uniquely identifies a consistent and reproducible benchmarking configuration}.
\end{principle}\hfill

Maintaining consistency in the benchmarks helps ensure performance comparisons of submitted simulators are more reasonable and fair. Nevertheless, as the benchmark suite is open to public submissions, we cannot guarantee perfect parity across all submitted simulators with regard to implementation decisions such as its choice of programming language or the extent to which it utilises multithreaded or GPU computation. Consequently, such context about each simulator is explicitly tracked in its metadata and may be filtered accordingly when comparing results. This allows, for instance, one to easily view only Pythonic simulators or to filter out GPU-accelerated simulators when visualising and comparing results.

Additionally, the consistency of the benchmark setup outlined in Principle~\ref{princ:consistency} in turn helps to provide consistency in the measured performance results of the submitted simulators. In other words, maintaining \textit{consistency} helps also to ensure \textit{reproducibility}. This is outlined in Principle~\ref{princ:reproduc}.

\hfill\begin{principle}[Reproducibility]
\label{princ:reproduc}
To help ensure the results measured by ABSTRACTS (and perhaps quoted on future publications which cite the project) on the performance of its simulators are reproducible, the following points are adhered to:
\begin{itemize}
    \item ABSTRACTS' version history is thoroughly maintained via GitHub.
    \item Every ABSTRACTS version maintains consistency according to Principle~\ref{princ:consistency}.
    \item The details of the benchmarking configuration of each ABSTRACTS version is explicitly recorded as per Principle~\ref{princ:consistency}.
    \item Every circuit and simulator is accompanied with explicit associated metadata.
    \item Every simulator defines its own environment, including explicit version numbers of any dependency libraries.
    \item Simulators with inherent randomness are encouraged to use a fixed seed to ensure deterministic computation.
\end{itemize}
\end{principle}\hfill

One reason (among many) why Principles~\ref{princ:consistency} and~\ref{princ:reproduc} are important is because we don't claim to have benchmarked all existing simulators (much less all potential future simulators) and as such we expect, and encourage, researchers and developers to submit their simulators to ABSTRACTS for benchmarking. For this reason, it is vital that consistency and reproducibility are preserved \textit{over time} such that the performance of new simulator submissions can be fairly compared against older ones benchmarked potentially months or even years prior, underpinning Principle~\ref{princ:extensibility}.

\hfill\begin{principle}[Extensibility]
\label{princ:extensibility}
The aim of ABSTRACTS is not to provide a snapshot in time of classical simulators and their relative performances but rather to provide a go-to resource for benchmarking and comparing (both present and) future classical simulators. Thus, we encourage new submissions and anticipate the set of simulators and circuits included in the project will grow over time.

To this end, we aim for strong extensibility by ensuring the following:
\begin{itemize}
    \item New simulators can be easily and readily submitted for benchmarking.
    \item New circuits or families thereof can likewise be easily and readily submitted for inclusion in the benchmark suite.
    \item The project is open to the public, accepting any reasonable submissions with minimal review.
    \item The specification for submitting simulators and circuits is very minimal and indiscriminate of programming language, simulation technique, etc.
    \item The repository itself is public, meaning anyone is free to download the included circuits and/or simulators for experimentation and testing of their own methods locally.
    \item The use of ABSTRACTS, including submissions and benchmarking on the official virtual hardware, is entirely free for the end user.
    \item The project welcomes more fundamental contributions to the repository (beyond submitting new simulators and circuits), where appropriate.
    \item The ABSTRACTS repo, including the circuits, results, and any figures produced via the dashboard, is licensed under the permissive MIT licence, allowing users to freely use, modify, and redistribute the software.
\end{itemize}
\end{principle}\hfill

Lastly, it is important that the project and the results it produces can be trusted as honest and faithful accounts of the performances of its simulators. In support of this, ABSTRACTS adheres to a principle of transparency, outlined in Principle~\ref{princ:transparency}.

\hfill\begin{principle}[Transparency]
\label{princ:transparency}
To help ensure ABSTRACTS and its results can be trusted, the following points are followed:
\begin{itemize}
    \item The project is open-source.
    \item Raw results are saved and publicly available.
    \item The project is well-documented and preserves version history, as per Principle~\ref{princ:reproduc}.
    \item Details of the cloud compute instance are explicitly documented.
    \item Thorough metadata is included throughout, including those associated with every benchmarking instance.
    \item Submitted simulators undergo basic review checks to ensure incorrect, malicious, or cheating submissions are not accepted.
    \item Anyone is free to raise potential issues publicly on the GitHub repository.
\end{itemize}
\end{principle}\hfill

Together, these four principles serve to provide credibility to the project and the results it produces.

\section{Benchmark Suite and Dashboard}

\subsection{Benchmark suite}

While the framework is designed to be extensible and we expect the set of included circuit classes to grow, we nevertheless provide a broad range of classes available in the initial (v0.1) release. These are summarised in Table~\ref{tab:classes}.

We chose these particular circuit classes because they have each been used for benchmarking classical simulators in previous papers \cite{kissinger2022simulating,kissinger2022sim2,ahmad-thesis,wira-paper,codsi2022classically} and they exhibit a range of different structures (or lack thereof) and target a range of different exponential scalings. As such, while individual circuit classes among these might favour particular simulation approaches, taken together this set avoids bias towards any one technique.

\begin{table}[h]
    \centering
    \begin{tabularx}{\linewidth}{l X}
        \hline
        Circuit Class & Overview \\

        \hline Random Clifford+T & Provides a baseline for unstructured circuits, with controlled variability across depth, qubit count, T-count, connectivity, and other such metrics. \\

        \hline Clifford+Toffoli/CCZ & Provides more structured baseline circuits, with its Toffoli/CCZ gates decomposed into T-gates, maintaining similar variability across metrics. \\

        \hline Exponentiated Paulis & Models circuits structured as those which appear in Hamiltonian simulations and variational algorithms, though restricted to Clifford+T phases. \\

        \hline Modified Hidden Shift & A well established classical simulation benchmark class, containing both Clifford and non-Clifford components. \\

        \hline IQP & Circuits consisting of commuting diagonals, which can be structured as phase polynomials, and with controlled variability across connectivity. \\
        
    \end{tabularx}
    \caption{Circuit classes included as benchmarks in ABSTRACTS v0.1.}
    \label{tab:classes}
\end{table}

\subsection{Dashboard}

The aggregated results produced by ABSTRACTS can be downloaded or visualised directly in the ABSTRACTS `\textit{dashboard}' webpage at {\url{https://mjsutcliffe99.github.io/ABSTRACTS/}}. Here, users can view the data plotted as an interactive graph, selecting which metrics the x- and y-axes correspond to, as well as which simulators and circuit classes to include in the plot. Other features include options to toggle between a linear or logarithmic scale on each axis, as well as options to bulk toggle simulators based on tags such as its programming language. Additionally, highlighting a particular data point displays its details, including those of the corresponding circuit and simulator. Figure~\ref{fig:dashboard} shows an example screenshot of the dashboard.

The particular screenshot of Figure~\ref{fig:dashboard} shows the CPU time versus the PyZX-reduced T-count \cite{pyzx} of three simulators (namely `\textit{PyZX BSS}' \cite{kissinger2022simulating}, `\textit{PyZX Magic Cat}' \cite{kissinger2022sim2}, and `\textit{Quimb}' \cite{gray2018quimb}) in strongly simulating a variety of randomly generated Clifford+T circuits. (Further details on these simulators or this circuit class can be found amongst their respective metadata in the repo \cite{abstracts_repo}.) Consistent with what is observed in this plot, these two PyZX methods are based on stabiliser decomposition and thus scale exponentially with the T-count, whereas the Quimb simulator is based on tensor contraction and hence is essentially agnostic of the T-count (but scales exponentially with the `\textit{contraction width}', which is a function of the circuit's graph structure). Furthermore, between the two PyZX methods, the `\textit{Magic Cat}' variant achieves markedly better scaling, able to simulate circuits of higher T-counts at greater speeds.

Example plots generated via the ABSTRACTS dashboard are shown in Figure~\ref{fig:plots}. Specifically, these show the CPU times for simulating random Clifford+T circuits, of fixed qubit counts and fixed (reduced) T-counts respectively, with two different types of simulator. These plots demonstrate that, as expected, the stabiliser decomposition simulator scales exponentially with the T-count of the circuits, whereas the statevector simulator scales exponentially with the qubit count. Evidently, neither simulator is universally `better' than the other but rather the preferred practical choice depends on the use-case.

\section{Roadmap}

Ultimately, our aim is to benchmark a wide range of classical simulators with ABSTRACTS across a broad family of circuit classes to provide extensive data comparing their relative performances under various circuit constraints. As such, the most prominent objective for the future of ABSTRACTS is to continue to expand the set of simulators and circuits it includes. We invite the broader community to participate in this endeavour by contributing new simulators as they are developed, though even at present there exist a great number of simulators which we plan to include for benchmarking on ABSTRACTS in the foreseeable future. Table \ref{tab:simulators} lists an extensive, though not exhaustive, set of such simulators.

\begin{table}
    \centering
    \begin{tabularx}{\linewidth}{>{\hsize=0.6\hsize}X c >{\hsize=0.1\hsize}X >{\hsize=1.3\hsize}X c}
        \hline
        Author(s) & Ref & Type & Overview & Status \\

        \hline
        Gray, 2018
        & \cite{gray2018quimb,gray2021hyper}
        & TC
        & Tensor network library supporting strong simulation via optimised tensor contraction
        & \tick \\
        
        \hline
        Kissinger \& van de Wetering, 2021
        & \cite{kissinger2022simulating,bss}
        & SD
        & Recursive BSS stabiliser decomposition \cite{bss} combined with ZX-calculus \cite{coecke2018picturing,wetering2020zx} simplification
        & \tick \\

        \hline
        Suzuki et al., 2021
        & \cite{suzuki2021qulacs}
        & SV
        & Statevector simulator for noisy variational quantum circuits
        & \cross \\
        
        \hline
        Kissinger et al., 2022
        & \cite{kissinger2022sim2}
        & SD
        & Stabiliser decomposition strategy based on $\ket{T}^{\otimes5}$ and $\ket{\text{cat}_n}$ decompositions
        & \tick \\

        \hline
        Vincent et al., 2022
        & \cite{vincent2022jet}
        & TC
        & A parallel tensor contraction engine for classical simulation
        & \cross \\

        \hline
        Codsi, 2022
        & \cite{Codsi2022Masters}
        & SD
        & Suite of new stabiliser decompositions, based largely on combined cat states and star states
        & \cross \\

        \hline
        Laakkonen, 2022
        & \cite{Laakkonen2022Masters}
        & SD
        & Suite of new stabiliser decompositions, based largely on graph states involving H-boxes
        & \cross \\

        \hline
        Codsi \& van de Wetering, 2022
        & \cite{codsi2022classically}
        & SD
        & Vertex-cutting based decomposition strategy targeting IQP circuits
        & \cross \\

        \hline
        Vinkhuijzen et al., 2023
        & \cite{vinkhuijzen2023limdd,vinkhuijzen2023efficient}
        & DD
        & Classical simulation technique based on decision diagrams
        & \cross \\

        \hline
        Cam \& Martiel, 2023
        & \cite{cam2023speeding}
        & TC
        & Improved tensor network contraction based on ZX-calculus
        & \cross \\

        \hline
        Koch, Yeung, \& Wang, 2023
        & \cite{koch2024contraction}
        & SD
        & Suite of new stabiliser decompositions, and corresponding strategy, based on `triangle' nodes
        & \cross \\

        \hline
        Javadi-Abhari et al., 2024
        & \cite{qiskit2024}
        & SV
        & Qiskit Aer: high performance state vector simulator
        & \tick \\

        \hline
        Sutcliffe \& Kissinger, 2024a
        & \cite{Sutcliffe_paramzx}
        & SD
        & GPU-accelerated pipeline for stabiliser decomp. simulation, based on a parametric ZX-calculus
        & \cross \\

        \hline
        Sutcliffe \& Kissinger, 2024b
        & \cite{Sutcliffe_procopt}
        & SD
        & Decomposition strategy based on weighing vertices on their potential for cascading vertex cuts
        & \cross \\

        \hline
        Sutcliffe, 2024
        & \cite{sutcliffe_kpar}
        & SD; TC
        & Hybrid strategy, using graph partitioning to combine stabiliser decomp. with tensor contraction
        & \cross \\

        \hline
        Koziell-Pipe, Yeung, \& Sutcliffe, 2024
        & \cite{koziell-pipe2024towards}
        & SD
        & Stabiliser decomp. strategy based on graph neural networks trained with reinforcement learning
        & \cross \\

        \hline
        Ahmad, 2024
        & \cite{ahmad-thesis,wira-paper}
        & SD
        & Suite of new stabiliser decompositions derived from ZX-rewriting and vertex cutting
        & \cross \\

        \hline
        Codsi \& Laakkonen, 2026
        & \cite{codsi2026unifyinggraphmeasuresstabilizer}
        & SD; TC
        & Hybrid approach, combining stabiliser decomp. and tensor contraction under a unified framework
        & \cross \\

        \hline
        Kuyanov \& Kissinger, 2026
        & \cite{kuyanov2026efficientclassicalsimulationlowrankwidth}
        & TC
        & Algorithm for contracting ZX-diagrams with rank-decompositions
        & \cross \\
        
    \end{tabularx}
    \caption{Notable methods and software for strong classical simulation suitable for benchmarking on ABSTRACTS. The simulation `types' include Stabiliser Decomposition (\textit{SD}), Tensor Contraction (\textit{TC}), State Vector (\textit{SV}), Decision Diagram (\textit{DD}), and hybrids thereof. Meanwhile, the `status' column indicates whether the method has yet been benchmarked on ABSTRACTS (as of v0.1).}
    \label{tab:simulators}
\end{table}

For many of these simulators, the source code is publicly available and may require only the work of coding a simple wrapper, though other cases may involve acquiring source code from generous authors or reimplementation from the theory or prototypes described in their related papers.

We also aim for future releases of ABSTRACTS to include a broader set of non-Clifford circuit classes, including some beyond the Clifford+T regime (such as those admitting arbitrary phases). Notably, these include, but are not limited to, the following:
\begin{itemize}
    \item Grover/Oracle circuits.
    \item Quantum Fourier Transform / Phase Estimation circuits.
    \item Variational / Quantum Approximate Optimisation Algorithm (QAOA) circuits.
\end{itemize}
This might involve importing established circuit suites from existing repositories, such as those for benchmarking circuit \textit{optimisers}, namely \textit{T-par} \cite{github-tpar,amy2014polynomial}, \textit{RevLib} \cite{wille2008revlib}, \textit{QASMBench} \cite{li2022qasmbench}, and \textit{MQT Bench} \cite{quetschlich2023mqtbench}.

Beyond including more simulators and circuits, future objectives for the ABSTRACTS project include the following:
\begin{itemize}
    \item Benchmarking \textit{weak} simulators.
    \item Inclusion of non-Clifford circuit classes beyond Clifford+T (e.g. arbitrary phases).
    \item Additional simulation tasks, such as computing marginal amplitudes.
    \item Simulation of circuits under different noise models.
    \item Improved hardware-resource profiling, including GPU and multicore CPU utilisation.
    \item More visualisation options in the dashboard, such as heatmaps and/or 3D plots.
    \item Automated measures, such as hidden random circuits, to detect cheating submissions.
\end{itemize}

As the project is intended to be rather open, we also welcome suggestions and contributions from members of the broader community.

\section*{Acknowledgements}

We acknowledge the use of large language models in assisting with the creation of YAML code for the GitHub Actions workflows, shell scripts for the benchmark runner and simulator wrappers, the Python script for aggregating the raw results data, and HTML code for the interactive dashboard. The use of LLMs in this capacity served as a time-saving resource for the non-scientific boilerplate code required in setting up the benchmark infrastructure and any LLM-generated code was carefully reviewed and verified. All else, including the writing of this manuscript, is entirely the work of the authors.

During the development of this project, we became aware of another team independently working on a similar project and with a comparable timeline to our own. As such, we would like to acknowledge the upcoming \textit{Quantum Gauntlet} project for benchmarking classical simulators and recognise the work of its team, namely Guillermo Alberto Perez and Matthias Lanzinger, and the discussions we shared with them.

\nocite{*}
\bibliographystyle{eptcs}
\bibliography{generic}
\end{document}